\documentclass[preprint]{elsarticle}
\usepackage[T1]{fontenc}
\usepackage[utf8]{inputenc}
\usepackage{array}
\usepackage{multirow}
\usepackage{amsmath}
\usepackage{amssymb}
\usepackage{graphicx}

\makeatletter

\pdfpageheight\paperheight
\pdfpagewidth\paperwidth

\makeatother

\usepackage{babel}

\journal{carbon}
\begin{document}
\begin{frontmatter}
\author{Benjamin N. Katz\corref{cor1}}
\cortext[cor1]{Corresponding author}
\ead{bnk120@psu.edu}
\author{Vincent Crespi}
\affiliation{organization={Department of Physics, The Pennsylvania State University, 104 Davey Lab, University Park, Pennsylvania 16802, USA}}
\ead{vhc2@psu.edu}
\date{\today}
\title{Healing of a Topological Scar: Coordination Defects in a Honeycomb Lattice}
\end{frontmatter}
\begin{abstract}
A crystal structure with a point defect typically returns to its ideal local structure within a few bond lengths of the defect; topological defects such as dislocations or disclinations also heal rapidly in this regard. Here we describe a simple point defect -- a two-fold atom incorporated at the growth edge of a honeycomb lattice -- whose healing may require a defect complex spanning many atoms. \textit{Topologically}, the two-fold atom disappears into a single ``long bond'' between its neighbors, thereby making a pentagonal disclination. But \textit{chemically}, this disclination occupies as much physical space as a six-fold ring. This incompatibility of chemistry and topology can cause a damped oscillation of the Gaussian curvature that creates an expansive healing region
, a topological scar.
\end{abstract}

\maketitle

\section{Introduction}
\label{Introduction} 

Defects in 2D materials~\citep{zou_defects_2017,komsa_physics_2022} can be classified by their physical~\citep{wu_spectroscopic_2017}, chemical~\citep{khossossi_recent_2020}, or topological character~\citep{yazyev_topological_2010,popov_designing_2017}. A wide range of defects such as Stone-Wales rotations, vacancies, grain boundaries, edge defects, substitutional defects, or chemisorbed species have been extensively studied in systems post-growth \citep{pantelides_defects_2012, banhart_structural_2010, boukhvalov_chemical_2008, hofer_direct_2019}, but less so during growth. Here we explore the kinetic consequences of a topological defect ---  a pentagonal disclination — that is created by a chemical impurity that \textit{changes the coordination number of one site on a hexagonal lattice}, and consider how the conflicting requirements of chemistry and topology around this \textbf{coordination defect} may be reconciled. 

We seek general phenomena that may apply across a wide range of 2D materials and impurities; thus we focus on the intrinsic geometrical character of the coordination defect, using graphene growth as an illustrative example. We find an unusual behavior --- to our knowledge not previously recognized ---  wherein a single two-coordinate impurity spawns a large defect complex as the nominal Gaussian curvature of the 2D sheet is \textit{overscreened} layer-by-layer while the defect is overgrown. For graphene specifically, this disruption in growth may trigger further non-idealities, such as an accumulation of sp$^3$-bonded material around the defect site. Recent results on improved graphene growth outcomes under the rigorous exclusion of oxygen\cite{amontree_reproducible_2024} are intriguing in this regard: although oxygen can have many roles in graphene growth (including substrate oxidation\citep{li_unexpected_2015, hu_roles_2017} and the modulation of nucleation density\citep{temiz_effect_2017, habib_review_2018, liang_exploring_2017, ani_critical_2018, hao_role_2013} or growth kinetics\citep{zhang_role_2021, habib_review_2018, srinivasan_oxygen-promoted_2018}) and post-growth (in delamination\cite{li_unexpected_2015, habib_review_2018, zhang_role_2021} or the visualization of defects\cite{saeed_chemical_2020}), the dramatic improvement in graphene growth rate, quality, and reproducibility under its exclusion and the accumulation of sp$^3$ carbon in its presence hints at a unique role for two-fold coordination defects in disrupting the bonding geometry of three-fold hexagonal network solids during growth. 

Atomistic simulations of graphene growth are challenging to equilibrate on computationally accessible timescales because they often become kinetically trapped in highly disordered structures\cite{chen_controlled_2022, gao_transition_2012}. For this reason, graphene growth is sometimes simulated using lattice Monte Carlo methods where the ideal hexagonal structure is enforced by fiat\cite{chen_kinetic_2019}.  Alternative Monte Carlo schemes allow the system to move off-lattice along a pre-specified slate of chemical reaction pathways\cite{whitesides_effect_2015}, through bond rotations that produce non-sixfold rings \cite{zhuang_evolution_2016}, or through vacancy motion informed by how the lattice deforms around defects \cite{trevethan_vacancy_2014}. Phase field methods can potentially resolve atomic positions and dislocation cores in grain boundaries \cite{taha_grain_2017}, while multiscale approaches may combine multiple methods \cite{esmaeilpour_multiscale_2023}. Considering the growth dynamics that we wish to capture, we follow a off-lattice Monte Carlo approach that can explore large deviations from six-fold geometries in a computationally efficient manner. To manage computational complexity, we constrain these simulations to sp$^2$ bonding geometries.

\section{Methods}
\label{methods}

Consider a two-coordinate atom substituted into a three-fold hexagonal honeycomb network, or equivalently a ligand that bonds tightly to one carbon atom on the growth edge, converting that carbon to being effectively two-coordinate. Topologically, the two-coordinate atom disappears into a ``long bond'' and the ring to which is it bonded becomes topologically five-fold, although it is chemically six-fold. For the system to grow around this coordination defect it must repair the topological error while avoiding large deviations in bond geometry, i.e.\ add one net heptagonal ring to remain asymptotically flat while also bridging the large gap created by the two-fold atom.  This generally requires more reach than is provided by a single seven-fold ring. 

\begin{figure*}[!t]
\includegraphics[width=6in]{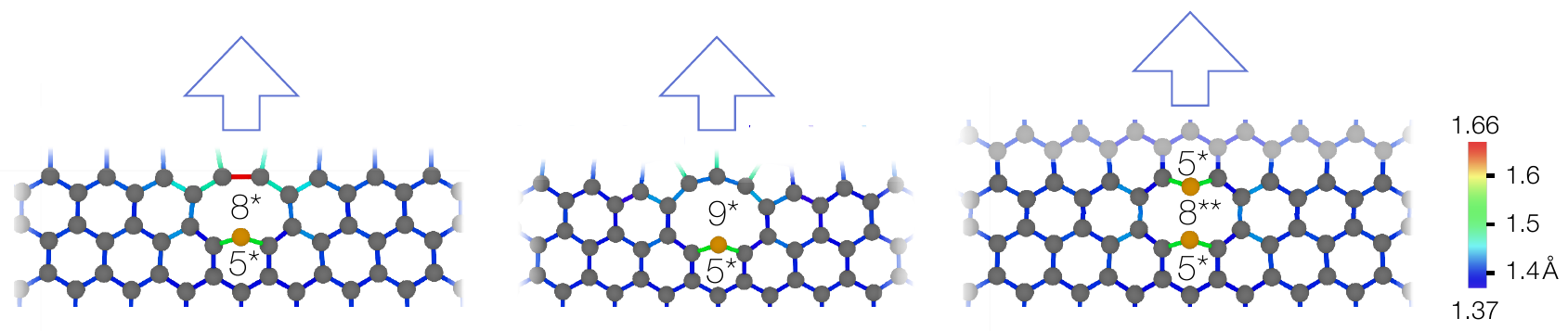}
\caption{Three possible resolutions to the growth beyond a single coordination defect (the two-fold orange atom): the first (which becomes the familiar single vacancy defect) and second produce an eight-fold or nine-fold ring respectively, while the third has two compensating coordination defects and resolves in a manner similar to a 5/8/5 divacancy structure. Bond lengths are color-coded as indicated in the scale bar at right.}
\label{possiblepaths}
\end{figure*}

There are several ways such a system could heal, with different numbers of atoms spanning the gap as depicted in Fig.~\ref{possiblepaths}. Attempting to heal with a seven-fold ring immediately above the coordination defect (as would happen in a simple dislocation) produces highly stretched bonds that break spontaneously during structural relaxation. Eight-fold and nine-fold rings reduce this bond strain but over-screen the Gaussian curvature of the coordination defect and so require further topological defects to achieve asymptotic planarity. If enough two-coordinate impurity atoms are available, then a nine-fold ring above the coordination defect could hold a second such atom and thus shrink topologically to an eight-fold ring, this two-coordinate atom then inducing a second topologically five-fold ring in the next layer to yield the structure shown in the right panel of Fig.~\ref{possiblepaths}. This geometry is akin to the well-known single or double vacancies in graphene and has been studied in the context of defects introduced into the interiors of already-grown sheets \citep{hofer_direct_2019}. However, if the impurity species is rare enough, then a second such atom may not become available before the system over-grows the initial defect. This could also occur by a doubly-coordinated unterminated (or hydrogen-terminated) carbon atom taking the place of the second defect atom: if not, this leaves the left and center panels of Fig.~\ref{possiblepaths} as possible routes for the overgrowth and healing of the coordination defect. 

Before moving further, we check if oxygen incorporation on the edge of graphene is favored over attachment to the basal plane to form an epoxide, since oxygen is a prime candidate for a coordination defect in graphene. To this end we create two comparison systems that contain the bonding geometries at issue without differences in the number of dangling carbon bonds. One is a 72-atom graphene unit cell with two oxygen atoms bound as epoxides and the other is the same sheet with a divacancy that is terminated with two oxygen atoms (effectively replacing a carbon dimer with oxygen atoms). Taking the chemical potential of carbon to be the energy of an atom in a 72-atom bulk graphene layer, density functional theory calculations performed in VASP \citep{kresse_ab_1993, kresse_efficient_1996, kresse_efficiency_1996, kresse_ultrasoft_1999} with the PBE functional on a $3\times 3\times 1$ k-point grid indicate a preference for oxygen to bind at a graphene edge by 0.94 eV over forming an epoxide. While this procedure compares defects that are internal to a graphene sheet, the geometries are similar to those of edge incorporation versus epoxide. We also note that oxygen is predicted to be favorable to bond with the edge of graphene, even if it has already attached to the copper substrate \citep{ma_oxygen_2015}, and can incorporate into other 2D materials during synthesis \citep{ding_oxygen-assisted_2023}. In light of the particularities of oxygen, we emphasize that it is only one possible choice for a two-coordinate species, and our main focus is on the more general topologically driven implications of under-coordination at the growth edge.

\section{Calculation}
\label{calculation}

For simplicity, we model growth on a graphene substrate, allowing the growing sheet to relax freely off-plane (because non-zero Gaussian curvature can induce such distortions) but keeping the substrate fixed. The metallic substrates often used experimentally can considerably complicate the edge energetics of \textit{unterminated} graphene sheets through substrate interactions with the sheet edge \citep{zhang_role_2014}, so we consider a regime of chemical potential in which the graphene edge is hydrogen-terminated, which weakens this interaction. In any case, many of the reconstructions that may occur at the growth edge for metallic substrates~\citep{wang_formation_2013,lee_copper-vapor-assisted_2018} are already considered in the branched growth model described below. 

Fig.~\ref{possiblepaths} shows our three starting points following incorporation of a two-fold coordination defect in monolayer graphene: an eightfold ring $(5^*8^*)$, a ninefold ring $(5^*9^*)$, or an eight-fold ring with a second two-coordinate atom $(5^*8^{**})$. An asterisk indicates a two-coordinate atom within that ring and the  integer just before gives the \textit{topological} size of the ring, not the number of atoms within it. As noted above, a 7-fold ring immediately above the coordination defect is not stable; a 10-fold ring is also disfavored relative to the 9-fold on energetic grounds. For convenience, we model the two-coordinate atom as a regular carbon atom with two hydrogens attached; this will automatically allow the $(5^*8^{**})$ pathway to be considered on an equal footing with the others through use of a hydrogen chemical potential. To verify generality with respect to the type of coordination defect we also implemented the defect as an oxygen atom and re-relaxed the most favored structures at each step described below (and also the full-row stages, favored or not); the essential results are unchanged (see Supporting Information).

\begin{figure*}[!t]
\includegraphics[width=6in]{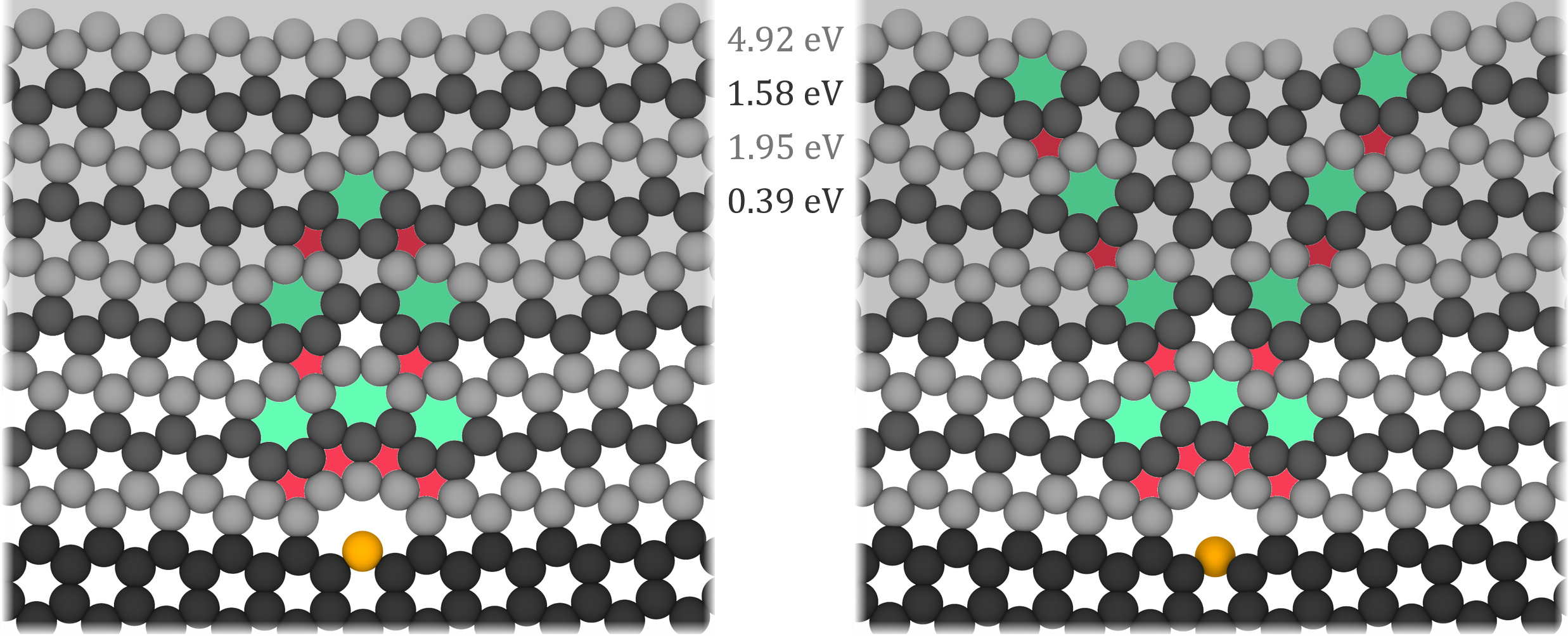}
\caption{Resolution of the growth pathway beginning as $(5^*9^*)(5^4)$. Successive rows are colored in alternating shades of gray, and non-six-fold rings are colored red (pentagons) and teal (heptagons).  The first three rows past the coordination defect (colored orange here) are added according to the algorithm described in the main text: later rows with gray background are added through energetic comparison of various completed rows. Numerical values at center give, row-by-row, the energy by which the structure of the right exceeds that of the structure on the left. In the right panel the defect complex spawns a grain boundary; in the left panel it forms a very large dislocation.}
\label{paths}
\end{figure*}

We assume step-flow (in 2D, kink-flow) growth down a zigzag edge, as done previously in modelling chemical vapor deposition on a weakly coupled substrate with low carbon flux\citep{luo_growth_2011}. Carbon atoms are added one at a time to flow the kink along the growth edge to form the successive rows of atoms depicted in alternating light or dark gray in Fig.~\ref{paths}. We assume that carbon atoms already three-fold carbon-coordinate within the bulk of the sheet do not undergo bond breaking and reformation during this process, but carbon atoms at the growth edge may reorganize in the vicinity of each successively added atom.

\begin{figure*}[!t]
\includegraphics[width=6in]{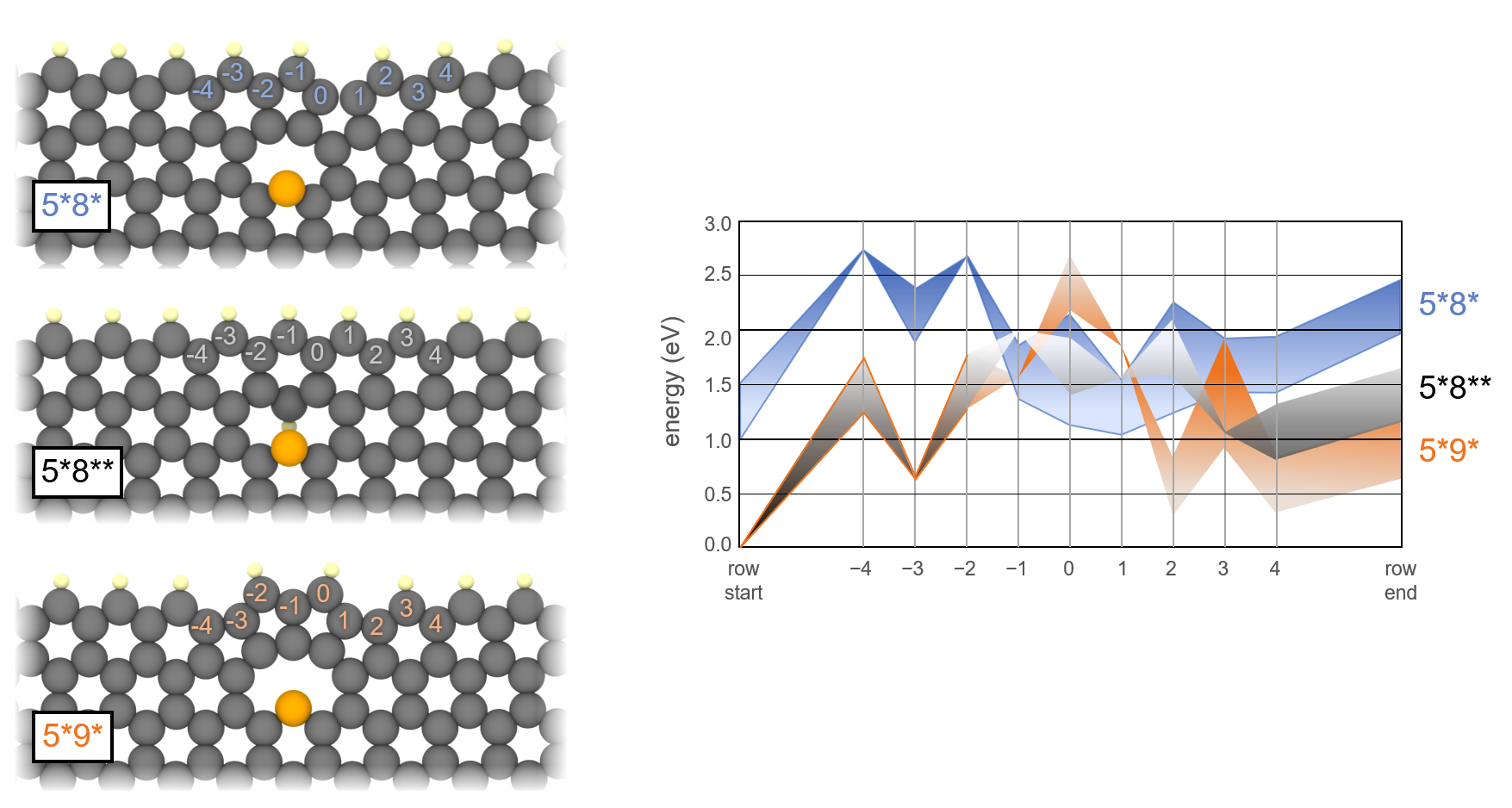}
\caption{Comparative step-by-step energetics for growth of two rows past a coordination defect, resulting in the three outcomes depicted at left, with the coordination defects represented by orange atoms. The atoms added are numbered in the sequence of growth (actual structures at these stages are given in Supporting Information). As the intermediate structures  differ in the number of hydrogen atoms, the graphs are broadened to cover a range of hydrogen partial pressures from $10^2$ to $10^0$ torr, fading towards the higher partial pressures (which are the lower hydrogen chemical potentials). The energies plotted are the energies of each growth pathway after the addition of the corresponding atoms.}
\label{energy-graph}
\end{figure*}

The initial $(5^*8^*)$, $(5^*9^*)$, or $(5^*8^{**})$ structures are hydrogen-terminated so that all atoms (other than the coordination defect) are three-coordinate. At this stage the $(5^*9^*)$ and $(5^*8^{**})$ pathways are identical; they are distinguished in the growth of the next row. We simulate the kink-flow growth of at least two additional layers across a branched tree of possible structural outcomes, terminating the highest-energy branches and allowing reconstructions of already-formed rings just behind the kink as it overgrows the coordination defect. We initiate the kink two six-fold rings to the side of the coordination defect, following the numbered atoms of Fig.~\ref{energy-graph}. Each added carbon atom can: (1) attach at the edge of the kink, becoming one-coordinate (or completing a ring and becoming two-coordinate); (2) attach to a one-coordinate atom with both becoming two-coordinate and thus completing a ring; or (3) insert into either of the two most recently completed rings. One- or two-coordinate here refers to the number of carbon neighbors, ignoring hydrogen termination.

We permit the addition of non-six-fold rings so long as they do not worsen the net Gaussian curvature deviation of the prior layer. For example, we permit five-fold rings in the next layer of $(5^*9^*)$, but no rings larger than six-fold. We never allow four-fold or three-fold rings. If the new atom attaches at the edge of the kink, its attachment point is presumed to be the two-coordinate atom nearest the three-coordinate bulk.  After this atom is added, we additionally allow one carbon atom (and the associated hydrogen) to move between the two most recently completed rings, shrinking one and growing the other. Any carbon atoms with less than three bonds are then singly terminated with hydrogen and the resulting structure is structurally relaxed.  As these structures may have differing hydrogen counts, we balance them with a hydrogen chemical potential taken for growth at 1000 $^\circ$C and $10^2$ torr hydrogen partial pressure, which is on the high side of the range (roughly $10^0$ to $10^2$ torr) expected to produce hydrogen-terminated graphene in experiment \citep{zhang_role_2014} (see Supporting Information). This chemical potential disfavors double hydrogen termination. We thus assume single termination during growth, excepting the 2-coordinate defects added by fiat.

Any structure whose relaxed energy is more than 10 eV above the lowest-energy structure of that step is pruned from the tree, while the remaining branches accrete additional carbon atoms as described above. This process continues until we complete the ring that is as far from the coordination defect at the one which began that row.  The row is then finished with six-fold rings (without reconstructions) for all remaining branches that are within 3.5 eV of the lowest-energy branch. The various outcomes may differ in carbon count as well as hydrogen count.  The carbon chemical potential is set to half the energy of a pair of carbon atoms that add another six-fold ring onto a kink on a pristine graphene edge. The branched tree does not reconstruct the interior of already-formed material (i.e.\ no Stone-Wales transformations or vacancy formation). By adding carbon atoms singly, it also captures almost all processes that may add carbon atoms in pairs, missing only ``dangling dimers''. 

Kink-flow growth across the defect is not the only way this system could grow. The defect might instead arrest the flow of the kink so that the graphene grows independently on each side of the coordination defect. Nevertheless, the material must eventually join across the defect. For arrested kink flow growth this would take the form of a grain boundary; a similar outcome is possible in full kink-flow growth, as shown by the right-hand structure of Figure~\ref{paths}. In both cases, the coordination defect spawns a semi-infinite defect complex. 

The energetics are modelled with the 2013 ReaxFF potential of Srinivasan et al. \citep{srinivasan_development_2015}, as implemented in the Large-scale Atomic/Molecular Massively Parallel Simulator (LAMMPS) \citep{plimpton_fast_1995}, which has been used to examine energetics and dynamics of highly defective carbon structures \citep{kroes_atom_2015, jensen_simulation_2015, li_reaxff_2018, sousa_mechanical_2018, roman_mechanical_2015, jensen_simulating_2018}. The coordination defect starts at the center of an edge of a graphene flake approximately 30 nm wide along the growth edge and 40 nm deep at the start of growth. It is supported on a fixed graphene substrate and singly terminated by hydrogen on the exposed edges: the substrate is aperiodic and extends for a buffer region of 60 to 80 nm (depending on the progression of growth) around the flake. The structural relaxation at each stage of the branched tree follows a multistep procedure that facilitates optimization of both fast and slow degrees of freedom. Relaxation begins with a 10 fs simulation using a bonded potential with distance-limited dynamics to remove any artifactual strain that may have been created upon introducing the new atoms. We then transition to the REAXX potential. A short series of conjugate-gradient relaxations is followed by a two-stage thermal anneal from 500 to 50 K and then from 50 to 0.1 K, each stage lasting 10 ps. This brief anneal facilitates the relaxation of slow long-wavelength degrees of freedom.  The terminating hydrogens buckle slightly out of the plane of the sheet and neither conjugate gradient nor damped dynamics can efficiently find their global energy minimum, creating a jitter of a few tenths of an eV across the entire system. Therefore two successive 50 ps damped-dynamics minimizations are followed by a final conjugate-gradient minimization, sampling the energy every 0.1 ps throughout this process; the lowest energy from this sampling is taken as the minimized energy of the structure at a $\sim$0.1 eV resolution.

\section{Results and Discussion}
\label{results}

Fig.~\ref{energy-graph} compares the energies of these three growth paths for the first row of growth, with images and energies of each stage in Supporting Information. Each free energy is given as a range, the upper and lower bounds of which are determined by the bounds on hydrogen partial pressure for the growth of hydrogen-terminated graphene at 1000 $^\circ$C, as discussed in Zhang et al.\citep{zhang_role_2014}. While the $(5^*9^*)(5^4)$ pathway (now extending the pathway names to include the ring structure of the next row) has an unusually large number of five-fold rings--- two more than required to retain planarity--- it nonetheless is the most favorable structure across most of the growth steps.  The two steps for which it is not the most favorable (numbered 0 and 1 in Fig.~\ref{energy-graph}) would require multiple ring-opening reconstructions behind or underneath the growth edge for it to switch to a different pathway. As the free energy gained in such a switch would be comparable to the formation energy of other common defects in graphene, this suggests that $(5^*9^*)(5^4)$ is a viable--- if not unique--- pathway. 

The $(5^*8^{**})(5^*)$ defect is fully healed (i.e.\ attains zero net gaussian curvature) after addition of a single row, while one of the two five-fold defects added in growth over the $(5^*8^*)(5^2)$ defect disappears upon addition of another row. These two structures should then grow further through the addition of six-fold rings, yielding a fairly small overall point defect complex. The $(5^*9^*)(5^4)$ defect, on the other hand, has over-screened the Gaussian curvature by two pentagons and so we propagate it for a third row, after which it has a one-heptagon excess (Fig.~\ref{paths}). As the curvature excess is now only a single ring, further growth is performed row-by-row, considering patterns of compensatory defect rings adjacent or near-adjacent to defect rings in the prior row, until the structure fully heals. This row-by-row method successfully reproduces the outcomes of the more detailed algorithm for all prior rows and is more computationally efficient. The lowest energy growth pathway, row-by-row, is depicted on the left side of Fig.~\ref{paths}; a close alternative is shown on the right. Although the total energies of these large defect complexes relative to a defect-free sheet is substantial, these two structures follow the lowest energy pathways conditional on the initial two-coordinate defect having formed and not immediately terminated in $(5^*8^{**})(5^*)$, i.e.\ they are kinetically favored outcomes for growth from an initial coordination defect nucleus.

A $(5^*8^{**})(5^*)$ termination whose second coordination defect is a carbon atom with two hydrogens attached is unlikely under low hydrogen partial pressure, and graphene can be grown under such conditions\citep{zhang_role_2014}. The larger defect complexes in our analysis tend to be more favored under such conditions (which also promote more kinetically controlled graphene growth). Coordination defects could be promoted by introducing during growth trace amounts of species that favor two-fold coordination to sp$^2$ carbon or species that favor one-fold coordination to carbon at the growth edge and bind tightly enough that they are not displaced during growth, since a `capped carbon'' is effectively a two-fold coordination defect.

The lower-energy pathway (the one on the left) takes \textit{four} further rows to heal, seven total rows beyond the original defect. The entire seven-row superstructure past the original $5^*$ coordination defect has fourteen non-sixfold rings and acts topologically as a single heptagon. This ``topological scar'' is thus a very large dislocation. The $(5^*$8$^{**})$ pathway grown to similar size would be much lower in energy, but that pathway cannot be easily recovered after the first row of $(5^*9^*)(5^4)$ has formed; the road to hell has been paved. This road may even fork into the pair of grain boundaries shown on the right-hand side of Fig.~\ref{paths}; this structure rises in relative energy under further growth as the two projecting grain boundaries extend. In this manner a two-coordinate impurity such as oxygen may nucleate grain boundaries in a honeycomb lattice (as seen in graphene for other impurities \citep{liu_situ_2014}). Note that our finite simulation size with free boundaries may favor the resolution of $(5^*9^*)(5^4)$ as a dislocation rather than a grain boundary; in a real system a grain boundary may be more compatible with the elastic relaxation of the sheet and the maintenance of favorable sheet/substrate interactions. 

\section{Conclusions}
\label{conclusions}

Although substitutional impurities occupy a single lattice site, we show how they can spawn large defect complexes if they introduce topological complications into growth beyond the defect.  More specifically, we describe how two-coordinate impurities in a honeycomb structure could be much more disruptive to growth than previously imagined, thus motivating their rigorous exclusion from the growth chamber when pursuing reproducible defect-free growth. The specific defect complexes that we describe have to our knowledge not been directly observed so far. In the case of graphene growth, it may be that the disruption caused by two-coordinate species carries beyond the constraints of our simulation to include formation of sp$^3$-bonded carbon. More broadly, a concerted effort to controllably introduce or exclude various under-coordinating species during 2D materials growth may provide a means to fine-tune sheet morphologies to either enhance growth of defect-free sheets or induce formation of more complex morphologies optimized for gas adsorption or catalysis.

\section*{Acknowledgements}
This work was supported by NSF awards NRT-DESE-1147985 and DMR-2011839 with primary support from the latter.
Visualisations were made with the free version of OVITO~\citep{stukowski_visualization_2009}.
Computations for this research were performed on the Pennsylvania State University's Institute for Computational and Data Sciences' Roar supercomputer.

\appendix

\bibliographystyle{elsarticle-harv}
\bibliography{./coordination_grain_boundaries}
\end{document}


\begin{center}
{\Large \bf Supporting Information}
\end{center}

\section{Chemical potential calculations}
\subsection{Hydrogen chemical potential}

We assume the hydrogen atoms in our system equilibrate to a reservoir of $\text{H}_2$ gas.  Following  methodology of \citep{walle_role_2002}, we use the following expression for the hydrogen chemical potential:
$$ -2\mu_\text{H}=E_{\text{H}_2}-k_BT\ln \left(Z_{\text{trans}}Z_{\text{rot}}Z_{\text{vib}}\right)=E_{\text{H}_2}+k_BT\ln\left(\frac{pV_Q}{k_BT}\right)-k_BT\ln Z_{\text{rot}}-k_BT\ln Z_{\text{vib}}$$
where we take the standard form for the translational partition function of an ideal gas.  
We calculate this at a temperature of 1000$^\circ$C and a pressure of $10^2$ torr, which are conditions within the expected regime for hydrogen-terminated graphene growth \citep{zhang_role_2014}, giving $k_BT\ln\left(\frac{pV_Q}{k_BT}\right)=-1.90\text{ eV}$. It is computationally intensive to account for the vibrational entropy of hydrogen bound to the sheet, which would be required to accurately compute the vibrational partition function; in addition, the vibrational excitations of $\text{H}_2$ are high-energy relative to $k_BT$ and thus not expected to contribute significantly to the chemical potential.  We thus neglect the $\ln Z_\text{vib}$ term.  The $Z_\text{rot}$ term becomes
$$Z_\text{rot}=\sum_{J=0}^\infty g_J e^{-\frac{E_\text{rot}(J^2+J)}{kT}}=\sum_{J=0}^\infty g_J e^{-\frac{J(J+1)\hbar^2}{2IkT}}$$
For $\text{H}_2$, $E_\text{rot}$, which is the energy of the lowest fundamental rotation mode, is 7.37 meV.  As we are not in the high-temperature limit, so we collect terms until they contribute less than a tenth to the value of $Z_\text{rot}$.  As $\text{H}_2$ rotation exchanges a single pair of fermions, the overall wavefunction needs to be antisymmetric.  Since H has nuclear spin 1/2, for each rotational state w/ quantum number $J$ there are $(I+1)(2I+1)$ symmetric functions and $I(2I+1)$ antisymmetric functions, which yields 3 symmetric and 1 antisymmetric function for $I=1/2$.  Functions with odd $J$ are already antisymmetric and even $J$ functions are symmetric, so odd $J$ needs symmetric spin and vice versa.  Thus $Z_\text{rot}$ splits into two functions, replacing $J$ by $2K$ and $2K+1$ and including the appropriate degeneracies:
$$Z_\text{rot}=\sum_{K=0}^\infty(4K+1)e^{-\frac{2(2K+1)E_\text{rot}}{kT}}\text{ for even }J,$$
$$Z_\text{rot}=\sum_{K=0}^\infty3(4K+3)e^{-\frac{(2K+1)(2K+2)E_\text{rot}}{kT}}\text{ for odd }J$$
and we collect terms in the overall sequence up to $J=10$ (so 6 terms in the first sequence and 5 in the second).  This gives us a value for $k_BT\ln(Z_\text{rot})$ of $0.361$ eV.

As we are simulating the structures in the ReaxFF potential in LAMMPS \citep{srinivasan_development_2015,plimpton_fast_1995}, we calculate the value of $E_{\text{H}_2}$ in this potential, and so our final value of $\mu_\text{H}$ comes to $\sim4.00$ eV, or $\sim92.13$ kcal/mol. In the main text we use a range when plotting the free energy along the various growth paths, corresponding to partial pressures from $10^0$ to  $10^2$ torr.

\clearpage
\section{Energy table for second layer growth}
\begin{table}[h]
\caption{\label{tab:energies} Table of compared energies (in eV) for the second layer of growth past the defect, as shown in Figure 3 in the main text.  These are the energies of the most favorable structure for each growth path at each step: multiple structures were evaluated at each step.}

\begin{ruledtabular}
\begin{tabular}{cccc}
Step  & $(5^*8^*)(5)$  & $(5^*9^*)(5^4)$  & $(5^*8^{**})(5*)$\tabularnewline
\hline 
row start  & $1.50$ ($-$1C,$-$1H) & 0.00 (+0C,+0H) & 0.00 (+0C,+0H)\tabularnewline
-4  & 2.74 (+8C,+0H)  & 1.76 (+9C,+1H) & 1.76 (+9C,+1H)\tabularnewline
-3  & 1.88  (+9C,$-$1H) & 0.62 (+10C,+0H) & 0.62 (+10C,+0H)\tabularnewline
-2  & 2.68 (+10C,+0H) & 1.78 (+11C,+1H) & 1.78 (+11C,+1H)\tabularnewline
-1  & 1.36 (+11C,$-$1H) & 1.56 (+12C,+0H) & 1.99 (+12C,+0H)\tabularnewline
0  & 1.13 (+12C,$-$1H) & 2.70 (+13C,+1H) & 1.93 (+13C,+1H)\tabularnewline
1  & 1.03 (+13C,$-$1H) & 1.86 (+14C,+0H) & 1.57 (+14C,+0H)\tabularnewline
2  & 1.23 (+14C,$-$2H) & 0.29 (+15C,$-$1H) & 2.10 (+15C,+1H)\tabularnewline
3  & 1.42 (+15C,$-$1H) & 0.92 (+16C,$-$2H) & 1.05 (+16C,+0H)\tabularnewline
4  & 1.42 (+16C,$-$1H) & 0.32 (+17C,$-$1H) & 1.31 (+17C,+1H) \tabularnewline
row end  & 1.95 (+17C,$-$1H) & 0.65 (+18C,$-$1H) & 1.65 (+17C,+1H) \tabularnewline
\end{tabular}
\end{ruledtabular}

Energies are set relative to the lowest energy of Step 0.  Step 0 is the previous completed row, then Step 1 begins with the addition of several carbons along the undefected region: each step beyond this adds an additional carbon atom (save for the last step, which completes the row): the energy given at each step is that of the lowest-energy structure found for the addition of that carbon.\end{table}

\clearpage
\section{Simulated structure positions and energies}
\subsection{Images of most favorable structures along growth paths}
\begin{figure*}[h]
\includegraphics[width=5in]{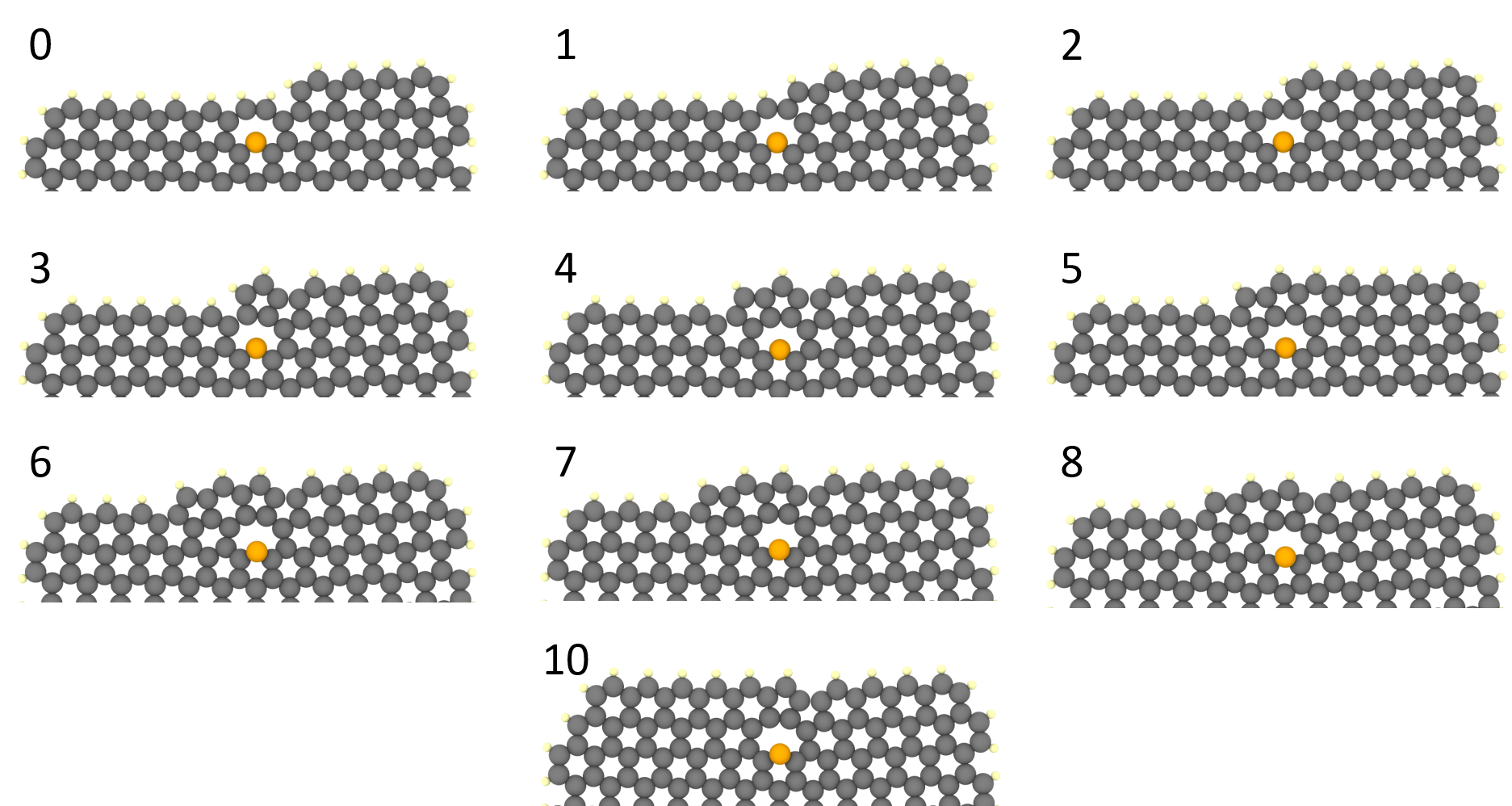}
\caption{The lowest-energy structure at each step for the $(5^*8^*)(5)$ growth pathway in the second layer.  Steps correspond to the table above; numbers correspond to the numbers in the position files and energy tables.}
\end{figure*}

\begin{figure*}[h]
\includegraphics[width=5in]{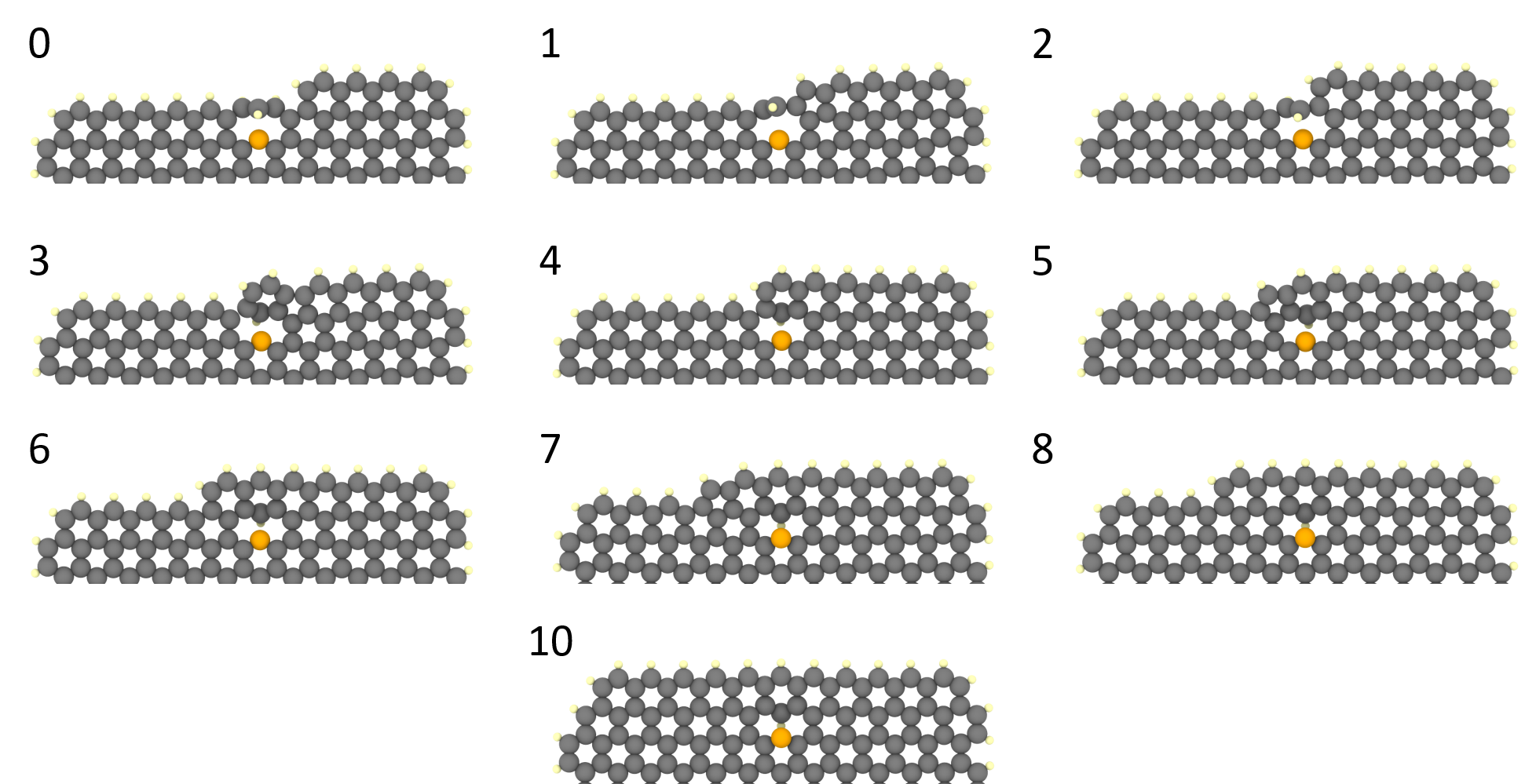}
\caption{The lowest-energy structure at each step for the $(5^*8^{**})(5^*)$ growth pathway in the second layer.  Steps correspond to the table above; numbers correspond to the numbers in the position files and energy tables.}
\end{figure*}

\begin{figure*}[h]
\includegraphics[width=5in]{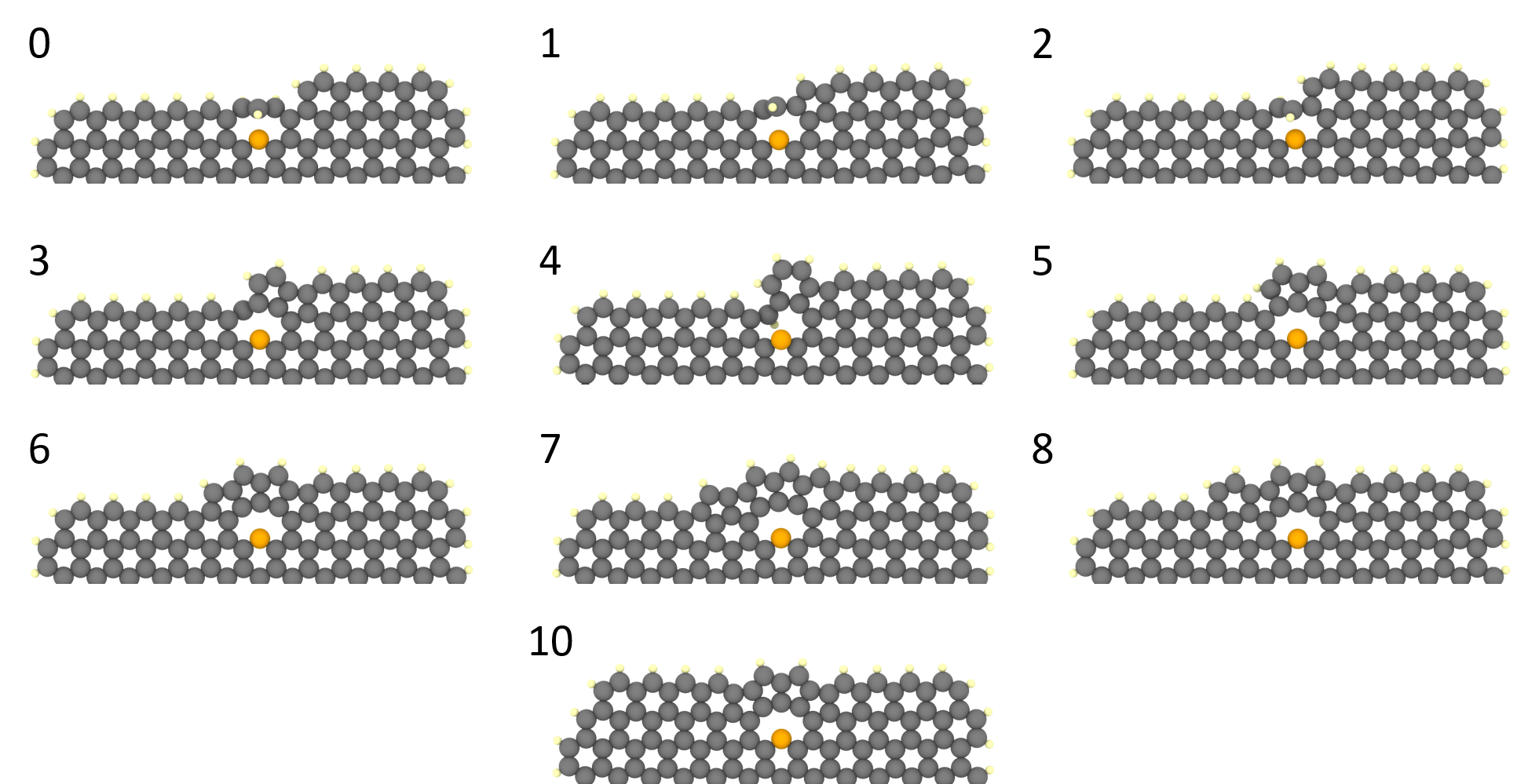}
\caption{The lowest-energy structure at each step for the $(5^*9^*)(5^4)$ growth pathway in the second layer.  Steps correspond to the table above; numbers correspond to the numbers in the position files and energy tables.}
\end{figure*}

\begin{figure*}[h]
\includegraphics[width=5in]{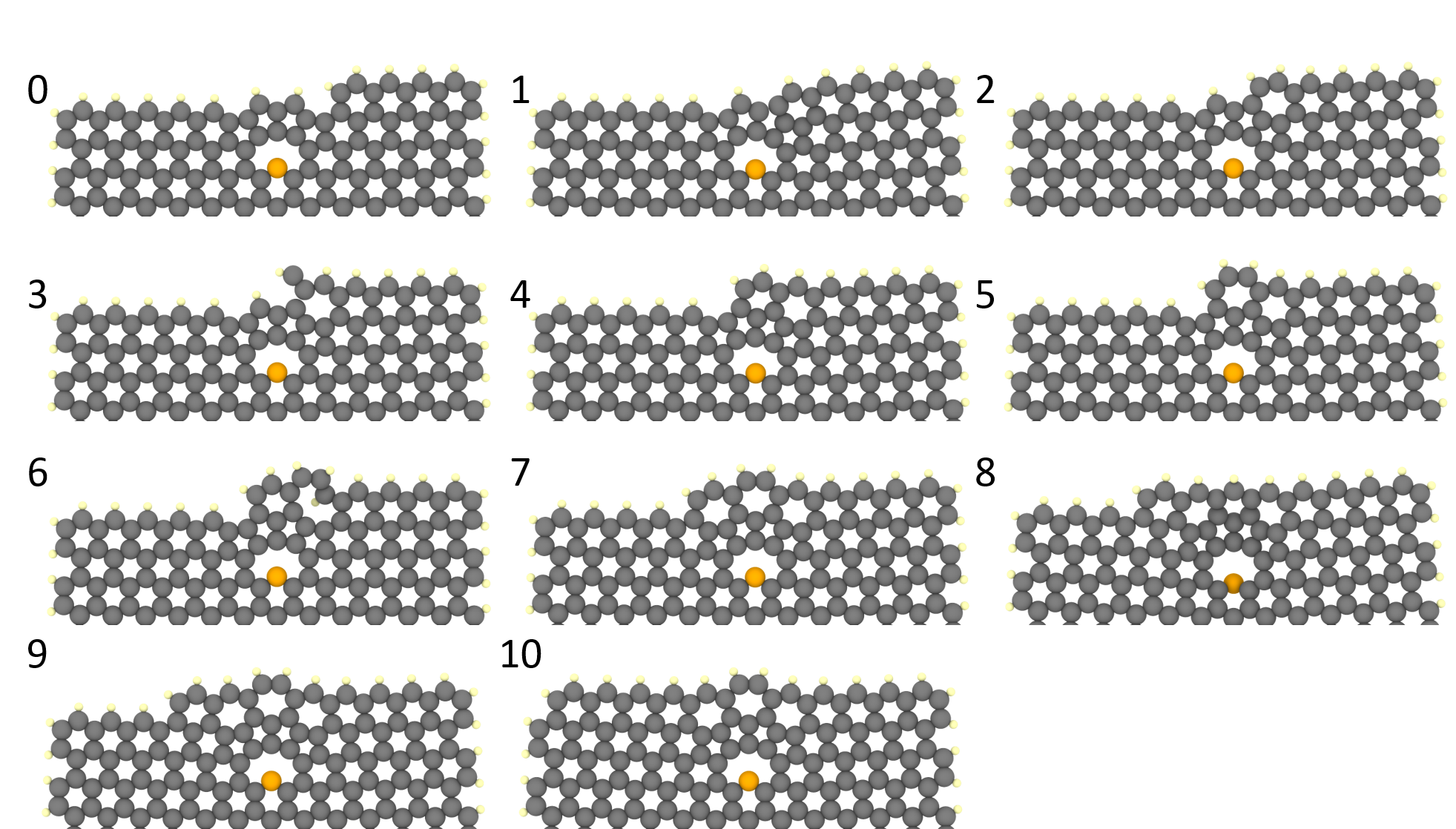}
\caption{The lowest-energy structure at each step for the $(5^*9^*)(5^4)$ growth pathway in the third layer.  Numbers correspond to the numbers in the position files and energy tables.}
\end{figure*}

\section{Guide to Structure Positions and Energies}
Simulated structure positions and energies are included in a ZIP file (`simulated\_positions.zip') as part of the Supporting Information.  The top-level directory features a folder for the structures in the $(5^*8^*)(5)$ growth path, divided into folders for the second and third rows, and a folder for the structures in the $(5^*8^{**})(5^*)$ and $(5^*9^*)(5^4)$ growth paths, divided into folders for the second, third, and succeeding rows.  The $(5^*8^{**})(5^*)$ and $(5^*9^*)(5^4)$ structures are commingled, as they share common structures along the growth pathway.  In the top-level directory is also provided a position file for the substrate, which (to save space) is omitted from the position files for all the structures, but was included in the simulation of the structures.

Files containing the energies, and a correspondence table to the structure filenames which identifies their ring counts are also provided, both in a CSV format broken apart into separate files for the $(5^*8^*)(5)$ growth path, the $(5^*8^{**})(5^*)$ and $(5^*9^*)(5^4)$ growth paths for the second row, the $(5^*9^*)(5^4)$ growth path for the third row, and the $(5^*9^*)(5^4)$ growth path for the rows beyond the third, as well as in an Excel spreadsheet featuring all these growth paths collected (and with functioning formulas to flag viable structures based on their energies).  The first column in the table gives the step identifier, which is also the name of the corresponding position file.  The formation energies given are relative to the energy of the structure with solely the initiating defect present.  As it may be an aid to the reader to examine the data alongside a well-recognized structure, step 10.18 in the second row of the $(5^*8^*)(5)$ path is the much-studied single-vacancy defect in graphene.

Selected structures are also evaluated with an initiating defect comprised of oxygen as opposed to the $\text{CH}_2$ group used elsewhere in this article: the most favorable structure for each added carbon (as determined by using the $\text{CH}_2$ group), and every end-row state.  While this does shift energies (primarily in those more defective structures which allowed the $\text{CH}_2$ group to rotate towards the substrate), it does not affect the choice of most favorable structure per row.  These energies are also included in the same file format as above.

\appendix
\bibliography{./coordination_grain_boundaries}